\documentclass[preprint,showpacs,preprintnumbers,amsmath,amssymb]{revtex4}
\usepackage{graphicx}% Include figure files
\usepackage{dcolumn}% Align table columns on decimal point
\usepackage{bm}% bold math

\begin{document}

\centerline{\bf Precision and Casimir Force Measurements in Superfluid Helium }
\centerline{\bf and Vacuum with Silicon Nitride Membranes}
%\medskip
\centerline{Steve K. Lamoreaux}
\centerline{\it email: steve.lamoreaux@yale.edu}
%\medskip
\centerline{Yale University, Department of Physics, Box 208120, New Haven, CT 06520-8120}
\centerline{\bf Abstract}
The recent discovery that silicon nitride membranes can be used as extremely high Q mechanical resonators makes possible a number of novel experiments, which include improved long range vacuum Casimir force measurements, and measurments of the properties of liquid helium below the lambda point.  It is noted that in the thermal correction to the Casimir force, the phase velocity of the excitations does not enter, with the force per unit area between parallel plates depending only on the temperature and distance between the plates.  Thus it appears as possible to measure the phonon analog of the finite temperature Casimir force in liquid helium.

\noindent
{\bf 1. Introduction}

The recent discovery that silicon nitride membranes can be used as high-$Q$ mechanical resonators opens possibilities for many new high sensitivity and high precision measurements $[1]$.  These membranes are formed by depositing a thin layer of silicon nitride on a silicon wafer, then etching away part of the silicon, leaving a free silicon nitride membrane, supported around its edges by the remaining silicon, as shown in $[1]$.  These supported film membranes, intended as transmission electron microscope substrates or X-ray windows, are commercially available $[2]$ with window dimensions up to 5 mm by 5 mm, and membrane thickness between 20 and 200 nm.  The X-ray windows can withstand atmospheric pressure, giving an indication of the membrane strength.

%\begin{figure}[h]
%\begin{center}
%\includegraphics[width=3.5in ] {membrane.eps} \caption{(a) Photo of a 1 mm by 1 mm by 50 nm Norcada x- ray membrane. (b) SEM of a 1 mm by 1 %mm by 50 nm
%membrane manufactured by SPI (graphic is from $[1]$).}
%\end{center}
%\end{figure}
Initial mechanical oscillation measurements on a 1 mm by 1 mm, 50 nm thick specimens indicate a $Q\approx 10^6$ at a temperature of 300 K, increasing to $Q\approx 10^7$ at 300 mK,  with a resonant frequency of about $\omega_0/2\pi=10^5$ Hz. These measurements were performed by observing, using fiber-coupled laser interferometry, the ring-down time after the membrane vibration was excited with a piezoelectric transducer.  The maximum amplitude before non-linearity becomes significant is about 0.18 nm.

The thermal noise limit to a force measurement is given by
\begin{equation}
S_f=\sqrt{4 k k_b T\over \omega_0 Q}\ \ \ {\rm N/\sqrt{Hz}}
\end{equation}
where $k\approx 30$ N/m is the effective spring constant of the membrane, $k_b$ is Boltzmann's constant and $T$ is the temperature,.  At 300 K, using the measured parameters, the thermal limited force sensitivities are
$ S_F=7\times 10^{-16}\ \ {\rm N/\sqrt{Hz}}$
while at 300 mK,
$S_f= 7 \times 10^{-18}\ \ {\rm N/\sqrt{Hz}}.$

Although these levels of sensitivity can be achieved with a micro- or nano-cantilever system, the advantage of the silicon nitride membranes is that the sensitive area can approach 25 (mm)$^2$.  In addition, the membranes are flat to order 1 nm over the entire area, and have a roughness less than 0.1 nm rms over the surface.
To illustrate the experimental advantages of these membranes, consider a measurement of the Casimir force between metal coated surfaces.  At long distances, the finite temperature correction dominates the force. If a probe with a spherical tip of radius of curvature $R$ is brought a distance $d$ from a flat surface, the force is
\begin{equation}
F= {1.2 Rk_bT\over d^2}.
\end{equation}
Taking $R=1$ cm, the signal to noise (per unit bandwidth), $S/N=F/S_F$ is unity when
$d= 26\ \ \mu{\rm m}$,
which is an order of magnitude larger distance for $S/N=1$ in experiments employing a torsion pendulum $[3]$.  For micro-cantilever measurements, the $S/N$ is about unity at $0.5 \ \mu$m (at this distance, the force is approximately proportional to $1/d^3$ for the sphere-plane geometry). (The conducting region on the membrane must have a diameter larger than 2 mm for the usual sphere-plane approximation to be valid.)  Measuring the Casimir force between metal films at such a large distance would resolve a theoretical controversy surrounding the thermal correction to this force.  In this instance the flatness and smoothness of the membrane are also of importance.

\noindent
{\bf 2. Research Projects in Fluids}

\noindent
Given that the study of the mechanical dynamics of silicon nitride membranes has begun only recently, our proposed research project discussion is divided into two sections.  First, we propose a further study of the membranes themselves, followed by a series of experiments.  

{\bf A. Membrane Resonator Studies}

\noindent
There are several issues to be further studied in the fabrication, mounting, and motion detection of silicon nitride membranes.
First, it is necessary to have a metallic coating on the membranes when used in Casimir or other force measurement experiments.  The effect of a metal film on the $Q$ remains to be fully studied, but we envision a small diameter circle near the center of the membrane with thin leads, or a very thin continuous coating, leading to the silicon substrate supported area.  For Casimir force measurements, the metallic film needs to be optically thick, usually of order 50 nm.  For other applications, a film thick enough to ensure that static charges can be drained from the measurement region is all that is required.  Given that the area to be coated is a few mm in diameter, very simple masking techniques can be employed, and the films directly applied using thermal evaporation.

Second, the ultimate $Q$ remains to be determined.  The ratio of silicon support mass to effective membrane mass is about $10^5$ for Norcada x-ray windows.  If the system was mounted on a very dissipative support, the $Q$ would be limited to $10^5$.  In addition, distortions of the silicon support can also lead to dissipation, which suggests that the largest silicon support possible is desirable, but this needs to be experimentally tested.

The detection of the membrane movement has been based on fiber laser interferometric measurements.  The thermal fluctuation surface displacement has a root-mean-square amplitude of 0.01 nm, which is easily optically detected.  This is also in the realm where the membrane motion could be detected electrically, and this offers some advantages for measurements in closed cryostats, for example. A small plate could be brought close to the (metallized) back surface of the membrane.  If the separation is 0.1 mm, the change in the capacitance for a .01 nm displacement is $10^{-5}$, so with 1 volt applied to the capacitor, it is possible to have a signal amplitude near 1 $\mu$V due to thermal motion at 300 K, which can be easily detected.  At 300 mK, the fluctuations are about 100 times smaller, and achieving this level of sensitivity either optically or electrically will be more challenging.

It remains to be determined whether a change in resonant frequency due to the gradient of the force, or a displacement of the membrane due to attraction, is the more effective measurement technique.

Finally, for some measurements, the area should be larger than the commercially available maximum of 5 mm$^2$.  Films could be either custom manufactured, or silicon nitride coated wafers could be purchased and etched which is a simple process for such large structures.

{\bf B. Measurement Application: Casimir force}

\noindent
As discussed previously, the Casmir force could be measured, using silicon nitride membrane, with unprecedented accuracy to comparatively very large distances.  These measurements will require at least a very thin conducting film to eliminate static charge effects, and to obtain the conducting metal Casimir force, the conducting film needs to be at least 50 nm thick.  However, the effect of thinner film can be accurately calculated, particularly in the long distance region ($d>10\ \mu$m).  At these large separations, the flatness and smoothness of the films is not critical, so the 1-2 nm flatness and 0.1 nm roughness, or film electric path effects, should not cause significant errors.  Another feature of the silicon nitride membrane is that the plane of the membrane is determined by the silicon substrate, and thus can be well-defined.  The most critical parameter in high accuracy measurement of the Casimir force is the absolute separation between the surfaces, and having the movement associated with the force measurement constrained to distances less than 0.1 nm will certainly be beneficial.

{\bf C. Measurement Application: Superfluid $^4$Helium Phonon damping and Casimir force}

\noindent At sufficiently low temperatures ($T<0.6$ K), the only
thermal excitations in superfluid helium are phonons. When the
phonons reflect from a moving membrane surface, momentum is
transferred to the phonons, causing the membrane motion to be
damped. The damping coefficient can be estimated for $T<0.1$ K,
\begin{equation}
\Gamma={2\pi^2(k_bT)^4\over \rho_{SN}\ t\ h^3 c_{ph}^4}
\end{equation}
where $\rho_{SN}$ is the silicon nitride mass density, $t$ is
the membrane thickness, and $h$ is Planck's constant, and
$c_{ph}=237$ m/s is the velocity of first sound in superfluid
He. This relationship is valid for membrane velocities
$v_m=a\omega_0 <10^{-4}\ {\rm m/s}$, where $a<0.2$ nm is the
amplitude,  that are much less than the Landau critical velocity
$v_L=60\ {\rm m/s}$.  The effective mass of the membrane might
be increased by the liquid helium, however, all damping effects
in the low velocity and low temperature are due solely to
existing phonons in the liquid, as described by Eq. (3).
Measurement of this damping time as a function of temperature
does not require a conducting coating if optical techniques are
used to detect the membrane motion.  The $Q$ due to phonon
damping is $10^{7}$ at about 30 mK.

For electromagnetic zero point energy, the Casimir force is
proportional to the speed of light ($c$). Applying the this
formalism to the phonon zero point field, the phonon Casimir
force is reduced by the ratio of the phonon velocity to the
speed of light, $c_{ph}/c\approx 10^{-6}$. However, the thermal
correction of the electromagnetic Casimir force, Eq. (2) above,
the speed of light does not enter, and by analogy, the thermal
Casimir force due to phonons in superfluid helium should be
one-half of Eq. (2) because there is only one (longitudinal)
mode for phonons. The electromagnetic Casimir force can be
suppressed by using a very thin conducting films on the
dielectric substrates, so the thermal phonon contribution to the
force can dominate at sufficiently large separations.  With the
increase in $Q$ at low temperature, measurement of the thermal
phonon Casimir force appears as possible.

{\bf D. Measurement Application: $^3$He concentration in superfluid $^4$He.}

\noindent
The most exciting prospective use of silicon nitride
membranes is the determination of $^3$He concentrations in
superfluid helium, an outstanding problem in low temperature
physics.  In particular, the ability to determine the
concentration at the part in $10^{-10}$ level is crucial for the
planned experiment to measure the neutron electric dipole moment
using ultracold neutrons and $^3$He stored together in
superfluid helium.  The membrane damping time for collisions
with $^3$He atoms is approximately
\begin{equation}
\Gamma={\pi 3^{3/2}\over 4} {\sqrt{ m_3^{*} k_bT} N_3 \over \rho_{SN} t}
\end{equation}
where $m_3^*=2.2 m_3$ is the effective $^3$He mass in
superfluid helium, and $N_3$ is the $^3$He number density.  

It should be noted that the damping time of a mechanical oscillator (quartz fiber)
provides the basis of the Langmuir vacuum gauge.$[4]$

The $Q$ due to this damping mechanism is of the same magnitude as
the intrinsic membrane $Q$ at $T=30$ mK for $N_3=10^{-10}N_4$,
at which temperature the phonon damping is relatively
negligible.  Thus, silicon nitride membranes can provide an
effective measurement for $^3$He concentrations above $10^{-10}$
mole fraction. Furthermore, the presence of $^3$He atoms can
give rise to a short-range attractive force between two surfaces
when the separations are comparable to the thermal wavelength of
a $^3$He atom, about 14 nm.  This force would have to be
discriminated from the phonon and electromagnetic Casimir
forces.
\vfill\eject
\centerline{\bf References}
\noindent $[1]$ B. M. Zwickl, W. E. Shanks, A. M. Jayich, C.
Yang,1 A.C. Bleszynski Jayich, J. D. Thompson, and J. G. E.
Harris, arXiv:0711.2263v1 [quant-ph]; Accepted for Publication,
Jour. Appl. Phys.

\noindent
$[2]$ See the products catalog at www.norcada.com, www.2spi.com, or www.tedpella.com.

\noindent
$[3]$ S.K. Lamoreaux, Phys. Rev. Lett. {\bf 78}, 5 (1997).

\noindent
$[4]$ I. Langmuir, J. Am. Chem. Soc. {\bf 35}, 107 (1913).  See also,
J. Strong et al., {\it Procedures in Experimental Physics} (Prentice-hall, New York, 1938) pp 146-148.

\end{document}